# Femtosecond Laser-induced Crystallization of Amorphous $Sb_2Te_3$ film and Coherent Phonon Spectroscopy Characterization and Optical Injection of Electron Spins

Simian Li,[1,3] Huan Huang,[2] Weiling Zhu,[1] Wenfang Wang,[1] Ke Chen,[1] Dao-Xin Yao,[1] Yang Wang,[2,a] Tianshu Lai,[1,a] Yiqun Wu,[2] and Fuxi Gan [2]

[1] State Key Laboratory of Optoelectronic Materials and Technology, School of Physics and Engineering, Sun Yat-Sen University, Guangzhou 510275, China

[2] Key Laboratory of High Power Laser Materials, Shanghai Institute of Optics and Fine Mechanics, Chinese Academy of Sciences, Shanghai 201800, China

[3] Hebei Key Laboratory of Optoelectronic Information and Geo-detection Technology，Shijiazhuang University of Economics, Shijiazhuang 050031, China

A femtosecond laser-irradiated crystallizing technique is tried to convert amorphous $Sb_2Te_3$ film into crystalline film. Sensitive coherent phonon spectroscopy (CPS) is used to monitor the crystallization of amorphous $Sb_2Te_3$ film at original irradiation site. The CPS reveals that vibration strength of two phonon modes that correspond to the characteristic phonon modes ($A_{1g}^1$ and $E_g$) of crystalline $Sb_2Te_3$, enhances with increasing laser irradiation fluence (LIF), showing the rise of the degree of crystallization with LIF and that femtosecond laser irradiation is a good post-treatment technique. Time-resolved circularly polarized pump-probe spectroscopy is used to investigate electron spin relaxation dynamics of the laser-induced crystallized $Sb_2Te_3$ film. Spin relaxation process indeed is observed, confirming the theoretical predictions on the validity of spin-dependent optical transition selection rule and the feasibility of transient spin-grating-based optical detection scheme of spin-plasmon collective modes in $Sb_2Te_3$-like topological insulators.

---

[a] Authors to whom correspondence should be addressed, E-mail: stslts@mail.sysu.edu.cn, ywang@siom.ac.cn

## I. Introduction

Crystalline $Sb_2Te_3$ film is an intriguing material due to their widely potential applicability in high efficient thermoelectricity, phase change storage and topological insulators. Therefore, its preparation and characterization becomes an active subject in material and condensed matter physics. Currently reported preparation methods of such a film mainly include metal-organic chemical-vapor deposition (MOCVD),[**1,2**] and molecular beam epitaxy (MBE).[3,4] In contrast, there are many methods reported to prepare amorphous or nanocrystalline $Sb_2Te_3$ films, such as microwave-assisted wet chemical route,[5] electrochemical deposition,[6] magnetron [7,8] or ion beam [9] sputtering, thermal evaporation,[10] and mechanical alloying [11] etc. These preparation methods of amorphous $Sb_2Te_3$ film have an advantage of inexpensive equipment over ones of crystalline $Sb_2Te_3$ film. Consequently, it is very satisfactory if amorphous $Sb_2Te_3$ film can be converted into crystalline one by some simple post-treatment, and hence very interesting to explore good post-treatment techniques. Thermal treatment or annealing has been reported as a post-treatment technique.[7,8] In this paper, femtosecond laser irradiation is tried as a new post-treatment technique to convert amorphous $Sb_2Te_3$ film into crystalline one. Meanwhile, the coherent phonon spectroscopy (CPS) is used to monitor the crystallization at the original crystallization site because CPS is very sensitive to microstructures,[1,2] and directly monitors characteristic phonon modes, while characteristic phonon modes are closely related to specific microstructures, as Forst et al [12] showed.

On the other hand, V-VI compounds, such as $Sb_2Te_3$, $Bi_2Te_3$ and $Bi_2Se_3$, were predicted to be topological insulators theoretically,[13] and topological insulator states have been observed experimentally in $Bi_2Te_3$ and $Sb_2Te_3$.[14,15] Topological insulator is a new phase of quantum matter which contains many new quantum physical phenomena. Recently, S. C. Zhang et al [16] predicted theoretically that surface plasmon mode will couple to spin wave due to the presence of long range Coulomb interaction, giving rise to hybridized "spin-plasmon" collective modes, and proposed an experimental scheme optically to detect the spin-plasmon collective modes. The scheme is based on the optical excitation and detection of transient spin grating which was already used for the measurement of electron spin transport in GaAs quantum wells.[17-19] However, the prerequisite of the scheme is the availability of optical injection of electron spins. Recently, Lu et al [20] predicted theoretically that spin-dependent optical transition selection rule is effective in $Sb_2Se_3$-like topological insulator thin film, showing that electron spins are allowed injection into the thin film by circularly polarized light. However, no experimental data reported yet. Here we will test the prediction in laser-irradiated crystallized $Sb_2Te_3$ thin film.

## II. Sample and Experiment Description

The $Sb_2Te_3$ film sample studied here is an as-deposited film of 10 nm thickness deposited on a glass substrate by RF magnetron sputtering using a stoichiometric target with a diameter of 60 mm. The base pressure in the deposition chamber is typically $4.0 \times 10^{-4}$ Pa. Sputtering is performed using Ar atmosphere at a pressure of 0.85 Pa. XRD shows the as-deposited thin film is in amorphous states. The femtosecond laser pulses from a Kerr lens mode-locked Ti: sapphire oscillator with a repetition rate of 94 MHz, a central wavelength at 860 nm and a duration of 60 fs, are directed into a standard pump-probe experimental setup with >10 ratio of pump to probe in intensity.[21] The pump and probe beams emerging from the pump-probe setup are transmitted through a convex lens of 75 mm focal length and focused to a same area on the sample located at the back focal plane of the lens. The differential transmission of the probe is detected by a photodiode and measured by a lock-in amplifier referenced at a modulation frequency of optical chopper which modulates the pump beam. In all experimental measurements, the pump fluence is kept constant at a low level of 0.039 mJ/cm$^2$ which does not lead to phase transition of the film, whereas different laser irradiation fluence is used to induce phase transition or crystallization of amorphous $Sb_2Se_3$ film.

## III. Experimental Results and Analysis

Fig. 1 shows the laser irradiation fluence (LIF) dependence of transient transmission change. Each transient trace is measured on a fresh spot which was irradiated for a few seconds by a femtosecond laser fluence indicated by the number over the transient. It is worth noting that the laser irradiation to a fresh spot is also made by the pump pulses by first increasing pump fluence to some higher level and irradiating a fresh spot for a few seconds, and then decreasing pump fluence back to a low level of 0.039 mJ/cm$^2$ used in the measurement of the irradiated spot. Some long-scale transient traces are also taken and plotted in Fig. 2. It is evident from Fig. 1 that an oscillatory component is superimposed on a conventional absorption saturation profile, which is just so-called coherent phonon oscillation induced by the displacive excitation.[22,23] There are several obvious features appeared in those transient traces. First, the oscillatory amplitude increases with LIF. Second, saturation amplitude (non-oscillatory or average component) decreases slowly with increasing LIF. Third, absorption saturation relaxes faster with the increase of LIF and evolves into absorption enhancement (negative signals) in a few picoseconds, as the transients show with the LIFs above 0.116 mJ/cm$^2$. Meanwhile, Fig.2 shows a long period oscillation occurred. The oscillatory amplitude increases with LIF, while the tail of the transient reduces and even becomes negative with increasing LIF. It is worth reminding that all measurements are made under a same low pump fluence of 0.039

mJ/cm$^2$ which does not result in phase transition. Therefore, all features mentioned above should originate from the different irradiation fluence. That is to say, laser irradiation has led to some irreversible change of amorphous Sb$_2$Te$_3$ films. We believe the change is just phase transition or crystallization of amorphous Sb$_2$Te$_3$ films. Based on LIF-induced crystallization of amorphous Sb$_2$Te$_3$ films, all features mentioned above can be explained well.

Laser irradiation is spatially non-uniform due to Gaussian profile distribution of a laser beam. The threshold fluence of crystallization will be reached first at the center of the irradiated spot, and hence LIF-induced crystallization will first occur at the center area of the irradiated spot. With increasing LIF, central crystallization area will become larger and the degree of crystallization enhanced. However, the area detected by the probe is always constant ~26 μm in diameter that is determined by the focusing lens, and hence is composed of central crystallization area and its surrounding amorphous area. Different LIF will result in different-size central crystallization area and different degree of crystallization. Experimental signature in Figs. 1 and 2 should reflect the increasing process of crystallization area and the degree of crystallization within the detection spot. Based on such a physical picture, the experimental phenomena shown in Figs. 1 and 2 can be explained very well.

The as-deposited film is amorphous, long-range disordered, phonon-free vibration and poor band structure. Pump fluence leads to only electron filling which contributes only to the saturation signature. On the other hand, crystallization area is long-range ordered, and has phonon vibration modes and good band structure. Pump fluence results in not only band filling-induced saturation signature, but also coherent phonon oscillation by displacive excitation [22,23] and band-gap renormalization (BGR) effect [24] which is easily observed due to ~0.1 eV narrow band gap of crystalline Sb$_2$Te$_3$.[13]

Our experimental signature should be the summation of both signals from amorphous and crystalline states within detection spot. At a low LIF of 0.039 mJ/cm$^2$, the transient trace presents no oscillation, as top curves shows in Figs. 1 and 2, suggesting a low LIF of 0.039 mJ/cm$^2$ can not lead to crystallization, and thus the transient signature reflects solely electron-filling-induced absorption saturation effect in amorphous $Sb_2Te_3$ film. With the increase of LIF, it can be found from Figs. 1 and 2 that oscillatory component appears and enhances. The high frequency oscillation in Fig. 1 is just from the coherent optical phonons,[1,2] whereas the low frequency oscillation in Fig. 2 is from coherent acoustic phonons [25] that provide the strong evidence for the occurrence of long-range order. The fact that the oscillatory amplitude becomes strong with increasing LIF, just reflects the increase of the degree of crystallization. In contrast, non-oscillatory saturation component decreases with increasing LIF, just reflecting enhancement of BGR effect,[24] and thus increase of the degree of crystallization and the formation of good band structure in the detection spot, as the fact shows that tails of the transient traces in Fig. 2 move down and even become negative with the increase of LIF. With the increase of LIF, the absorption saturation evolves into absorption enhancement or negative signal in a few picoseconds, which may be attributed to the cooperative effect of the enhancements of BGR and coherent acoustic phonon modulation.

To identify the excited optical phonon modes to understand LIF-induced phase transition, low-pass digital filter is used to obtain non-oscillatory components which are subtracted from the transients so that the oscillatory components are acquired and FFT-transformed, as shown in Fig. 3 (a) and (b), respectively. It is evident that a main peak occurs at 3.63 THz, while a weaker one at 1.99 THz. Both peaks enhance with increasing LIF, showing LIF dependence of phase transition. The weak peak at 1.99

THz may be assigned to $A_{1g}^1$ mode of crystalline Sb$_2$Te$_3$,[1,12] while the main peak at 3.63 THz attributed to E$_g$ mode [12] because these modes differ only slightly from the frequencies obtained in cw Raman experiment for crystalline Sb$_2$Te$_3$.[26] The $A_{1g}^1$ and E$_g$ modes are the characteristic optical phonon mode of crystalline Sb$_2$Te$_3$. Therefore, their enhancement with increasing LIF once again shows the increase of degree of crystallization of amorphous Sb$_2$Te$_3$ films, revealing that LIF indeed enable the crystallization of amorphous Sb$_2$Te$_3$ film. As a result, femtosecond laser irradiation is proven to be a good post-treatment technique for the conversion of amorphous Sb$_2$Te$_3$ film into crystalline film. If a large power femtosecond laser is used and focused to a larger spot or a longer line using a cylindrical lens, large area crystalline Sb$_2$Te$_3$ film may be prepared by scanning the amorphous Sb$_2$Te$_3$ film line by line in the film plane at an appropriate speed.

## IV. Test of Spin-dependent Optical Transition Selection Rule

Lu et al [20] predicted theoretically that spin-dependent optical transition selection rule is effective in Bi$_2$Se$_3$-like films. They showed that right-handed (σ$^+$) or left-handed (σ$^-$) circularly polarized light can excite spin-up or spin-down electrons in conduction band, respectively, which provides a theoretical support with the transient-spin-grating-based optical detection scheme of spin-plasmon collective modes proposed by Zhang et al.[16] However, no experimental evidence reported so far. Here we will test the theoretical prediction of Lu et al [20] on a femtosecond laser-irradiated crystallized Sb$_2$Te$_3$ film. Time-resolved circularly polarized pump-probe spectroscopy is used to study electron spin relaxation dynamics in the Sb$_2$Te$_3$ film. This spectroscopy is sensitive to electron spin polarization and has been used extensively to investigate electron spin relaxation dynamics in semiconductors.[21,27,28] A broadband achromatic 1/4-wave plate is inserted into pump and probe light paths to switch linearly polarized light to $\sigma^+$ or $\sigma^-$ circularly polarized light.[21] Differential transmission transient traces for co-helicity ($\sigma^+,\sigma^+$) and cross-helicity ($\sigma^+,\sigma^-$) circularly polarized pump and probe beams are taken in a crystallized $Sb_2Te_3$ film irradiated by a LIF of 0.173 mJ/cm$^2$ and plotted in Fig. 4. They are different apparently for ($\sigma^+,\sigma^+$) and ($\sigma^+,\sigma^-$) pump and probe beams. The oscillatory component still describes the vibration of optical phonon, being the same as that in Fig. 1. Non-oscillatory component transient traces are acquired by low-pass digital filtering and plotted in the inset of Fig. 4, and more clearly show the difference between ($\sigma^+,\sigma^+$) and ($\sigma^+,\sigma^-$) transient traces. The crossing of two transient traces near ~1 ps may be attributed to acoustic phonon modulation, as Fig. 2 shows. If there were no the modulation, the ($\sigma^+,\sigma^+$) trace would be always stronger than ($\sigma^+,\sigma^-$) one in present time scale, as shown in Refs. [21] and [28]. As the inset shows in Fig. 4, the decay of ($\sigma^+,\sigma^+$) transient trace reflects the decrease of spin-up populated electron density excited by $\sigma^+$ pump pulses due to spin relaxation and recombination, while the first rising and then decay of ($\sigma^+,\sigma^-$) transient trace does the initial increase of spin-down populated electron density due to electron spin relaxation from spin-up to spin-down states, and then decay due to recombination, respectively.[21] Meanwhile, a transient trace with linearly polarized pump and probe beams is also taken, but not shown in Fig. 4 for the clarity of the figure. It is electron spin independent and always located between the ($\sigma^+,\sigma^+$) and ($\sigma^+,\sigma^-$) transient traces, as usually shown in semiconductor spin dynamics,[21,27,28] which further confirms that the difference between ($\sigma^+,\sigma^+$) and ($\sigma^+,\sigma^-$) transient traces originates from electron spin relaxation. From the transient signal intensity I($\sigma^+,\sigma^+$) and I($\sigma^+,\sigma^-$) at t = 0 of ($\sigma^+,\sigma^+$) and ($\sigma^+,\sigma^-$) transient traces in the inset of Fig. 4, an optical injected initial degree of electron-spin polarization can be estimated by the formula $[I(\sigma^+,\sigma^+) - I(\sigma^+,\sigma^-)]/[I(\sigma^+,\sigma^+) + I(\sigma^+,\sigma^-)]$ as ~26%. Theoretically Lu et al found a $\sigma^+$ or $\sigma^-$ light couples to two Dirac hyperbolas inequivalently, creating spin-polarized electrons.[20] When the photon energy is equal to the gap $\Delta$, the $\sigma^+$ or $\sigma^-$ light couples only to one Dirac hyperbolas, causing highly spin-polarized electron injection at Dirac point. On contrary, as the photon energy is larger than the gap $\Delta$, the $\sigma^+$ or $\sigma^-$ light can couple to both Dirac hyperbolas, causing a lowly spin-polarized electron injection at a k point away from Dirac point. The degree of spin polarization of the photoexcited electrons can be calculated by the interband transition matrix element $\pi_{cv}^{\pm}(k,\tau_z) \propto (1 \pm \tau_z \cos\theta)$, where $\cos\theta = \Delta/[\varepsilon_c(k) - \varepsilon_v(k)]$, $\tau_z$=1 or -1 being the index of Dirac hyperbolas and + (-) corresponding to right- (left-) handed circularly polarized light, while $\varepsilon_c(k)$ and $\varepsilon_v(k)$ describe the dispersion of conduction and valance bands of Dirac fermions.[20] The gap $\Delta$ of $Sb_2Te_3$ is about 0.1~0.25 eV from the band structure calculation[13,29] and can be increased by the finite thickness of thin film.[20] The photon energy we used is 1.44 eV. From these data, we can estimate $\cos\theta \approx 0.07 - 0.17$ and the degree of spin polarization of the photoinjected electrons,

$$P(k) = \frac{|\pi_{cv}^{\pm}(k,\tau_z)|^2 - |\pi_{cv}^{\pm}(k,-\tau_z)|^2}{|\pi_{cv}^{\pm}(k,\tau_z)|^2 + |\pi_{cv}^{\pm}(k,-\tau_z)|^2} \approx 14 \sim 33\%.$$ The theoretical value is close to our experimental data. On the other hand, our photon energy is much larger than the gap $\Delta$, which can cause possible transitions between bulk conduction and valence bands besides surface state transition. From the band structure of bulk $Sb_2Te_3$ crystal calculated by Wang et al,[29] it is possible for the transitions between three lower conduction and two upper valence bands near high symmetric $\Gamma$ point to be excited by 1.44 eV photons, creating spin unpolarized or low spin-polarized electrons in

conduction band. This can decrease the spin polarization to some extent. However, since our sample is a very thin film, the contribution from the bulk may be suppressed and the surface states become dominant. Clearly our experiment provides evidence to optically injected electron-spin polarization, and thus confirms the prediction of Lu et al, showing the feasibility of optical detection scheme of spin-plasmon collective modes proposed by Zhang et al.[16] Meanwhile, it also shows a good quality of LIF-induced crystallized $Sb_2Te_3$ film.

## V. Conclusion

A new post-treatment technique for conversion of amorphous $Sb_2Te_3$ film into crystalline film, femtosecond laser irradiation, is developed and executed. Sensitive CPS is used to monitor LIF-induced crystallization of amorphous $Sb_2Te_3$ film at original irradiation site. It is found that the vibration strength of two phonon modes, corresponding to the characteristic phonon modes ($A_{1g}^1$ and $E_g$) of crystalline $Sb_2Te_3$, enhances with increasing LIF, showing the rise of the degree of crystallization with the increase of LIF within the experimental range of LIF. It is shown that good quality crystalline $Sb_2Te_3$ film can be prepared by this post-treatment technique. Time-resolved circularly polarized pump-probe spectroscopy is used to investigate electron spin relaxation dynamics of the crystallized $Sb_2Te_3$ film. Spin relaxation process indeed is observed, confirming the theoretical prediction on spin-dependent optical transition selection rule and the feasibility of transient spin-grating-based optical detection scheme of spin-plasmon collective modes in $Sb_2Te_3$-like topological insulators.

## Acknowledgements

This work is partially supported by National Natural Science Foundation of China under grant Nos. 50872139, 10874247, 61078027 and 11074310, National

Basic Research Program of China under grant Nos. 2007CB935402 and 2010CB923200, and Natural Science Foundation of Guangdong Province under grant No. 9151027501000077 as well as doctoral specialized fund of MOE of China under grant No. 20090171110005.

Fig.1 (Color online) Laser irradiation fluence dependence of coherent optical phonon oscillation dynamics. The horizontal dashed lines denote zero-baselines which are shifted upward for clarity. The numbers indicated over the curves stand for the laser irradiation fluence used to irradiate a fresh spot on amorphous $Sb_2Te_3$ film. Pump excitation fluence is kept constant at a low level of 0.039 mJ/cm$^2$ in all transient measurements.

Fig. 2 (Color online) Laser irradiation fluence dependence of coherent acoustic phonon oscillation dynamics. The horizontal dashed lines denote zero-baselines which are shifted upward for clarity. The numbers indicated over the curves stand for the laser irradiation fluence used to irradiate a fresh spot on amorphous $Sb_2Te_3$ film. Pump excitation fluence is kept constant at a low level of 0.039 mJ/cm$^2$ in all transient measurements.

Fig. 3 (a) The damping oscillatory decay of coherent phonons evolved from Fig. 1 with non-oscillatory components removed. The horizontal dashed lines denote zero-baselines which are upward for clarity. (b) The FFT spectra corresponding to the oscillatory traces in (a). Zero-baselines are also shifted upward for clarity.

Fig. 4 (Color online) Circularly polarized pump-probe transient differential transmission traces measured in the crystallized $Sb_2Te_3$ film irradiated by a laser irradiation fluence of 0.173 mJ/cm$^2$. The ($\sigma^+$,$\sigma^+$) and ($\sigma^+$,$\sigma^-$) denote the same and opposite circularly polarized helicities of pump and probe beams, respectively. The inset shows the non-oscillatory components of the transients.

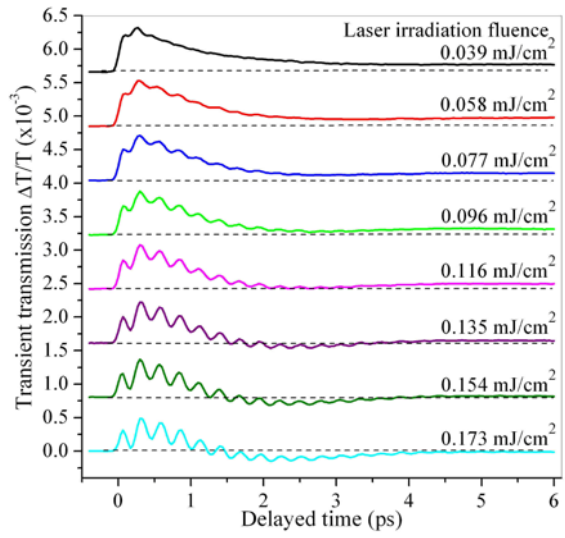

Figure 1

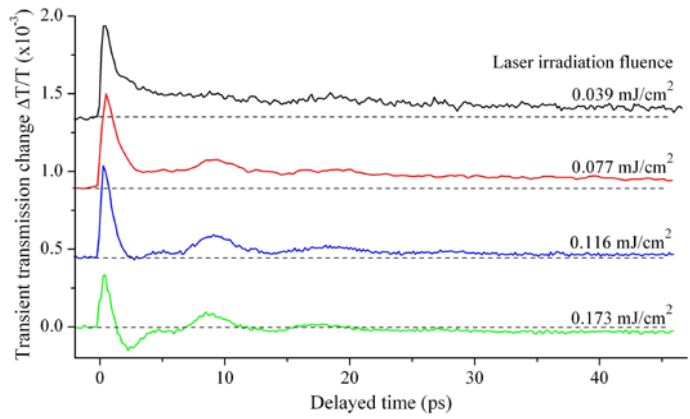

Figure 2

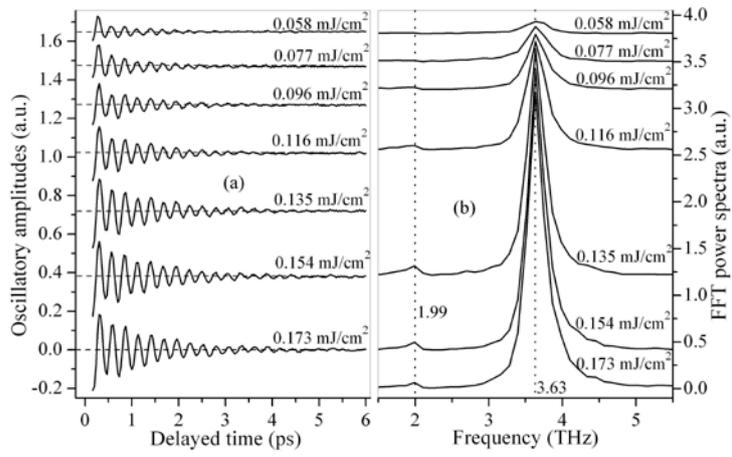

Figure 3

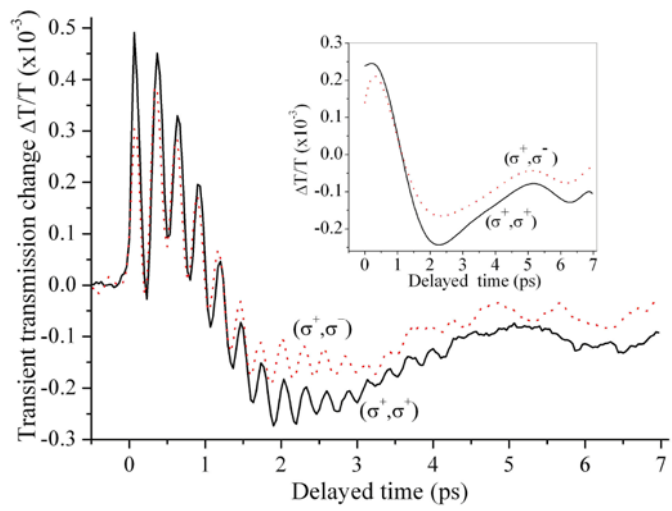

Figure 4